\documentclass[twocolumn,prl,showpacs,amsmath,amssymb]{revtex4}

\usepackage{graphicx}

\def\half{\frac{1}{2}}

\newcommand{\beq}{\begin{equation}}
\newcommand{\eeq}{\end{equation}}
\newcommand{\beqa}{\begin{eqnarray}}
\newcommand{\eeqa}{\end{eqnarray}}

\def\be{\begin{equation}}
\def\ee{\end{equation}}
\def\bea{\begin{eqnarray}}
\def\eea{\end{eqnarray}}
\def\bma{\begin{mathletters}}
\def\ema{\end{mathletters}}

\def\0{\overline{0}}

\def\q0{\underline{0}}

\def\one{\leavevmode\hbox{\small1\normalsize\kern-.33em1}}

\def\bra#1{\langle#1|} \def\ket#1{|#1\rangle}

\def\proj#1{\ket{#1}\!\bra{#1}}

\def\opone{\leavevmode\hbox{\small1\kern-3.8pt\normalsize1}}

\begin{document}

\title{From Bell's Theorem to Secure Quantum Key Distribution}

\author{Antonio Ac\'\i n$^{1}$, Nicolas Gisin$^2$ and Lluis Masanes$^3$}

\affiliation{ $^1$ICFO-Institut de Ci\`encies Fot\`oniques,
Mediterranean Technology Park, 08860 Castelldefels (Barcelona), Spain\\
$^2$GAP-Optique, University of Geneva, 20, Rue de
l'\'Ecole de M\'edecine, CH-1211 Geneva 4, Switzerland\\
$^3$School of Mathematics, University of Bristol, Bristol BS8 1TW,
United Kingdom
 }

\date{\today}

%%%%%%%%%%%% Abstract %%%%%%%%%%%%%%%%%%%%%%%%%%%

\begin{abstract}
Any Quantum Key Distribution (QKD) protocol consists first of
sequences of measurements that produce some correlation between
classical data. We show that these correlation data must violate
some Bell inequality in order to contain distillable secrecy, if
not they could be produced by quantum measurements performed on a
separable state of larger dimension. We introduce a new QKD
protocol and prove its security against any individual attack by
an adversary only limited by the no-signaling condition.
\end{abstract}


\maketitle

In 1991 Artur Ekert published his seminal paper {\it Quantum
Cryptography based on Bell's theorem} \cite{Ekert}. For the first
time, it was argued that quantum nonlocality could be good for
something! It was already known, since 1964, that quantum
correlations allow one to perform some tasks classically
impossible, like violating Bell's inequality between space-like
separated regions \cite{Bell}. But Ekert's proposal was the first
addressing a useful task and this had a huge impact on the
development of Quantum Information Science. Yet, none of today's
security proofs of Quantum Key Distribution (QKD) make a direct
use of quantum nonlocality (see however the recent work of Ref.
\cite{BHK}). The existing proofs are either based on the
no-cloning theorem, or on the {\it monogamy} (i.e. non
shareability) of entanglement. All proofs heavily exploit the
Hilbert space artillery of quantum physics.

In the recent years, quantum nonlocality underwent yet another
important twist. Thanks to the seminal paper by Popescu and
Rohrlich \cite{PR}, it was realized that one can study {\it
nonlocality without Hilbert space}. Indeed, although quantum
states violate Bell inequalities, there is nothing quantum in the
derivation of a Bell inequality. The picture is richer when one
adds the assumption of no-signaling, i.e. that the correlations
between distant partners cannot be used to send information, as is
the case for quantum correlations. The no-signaling principle
suffices to severely limit the set of correlations. Formally, a
correlation is a conditional probability distribution
$P(a,b|x,y)$, where $a$ and $b$ are Alice and Bob's output data,
respectively, and $x$ and $y$ are their choices of inputs. For
instance, $x$ and $y$ could be their choice of measurement
settings and $a$ and $b$ the obtained results. The no-signaling
condition implies that the marginals are independent of the
partner's input:
\begin{equation}\label{nsprinciple}
    P(a|x,y)=\sum_bP(a,b|x,y)=P(a|x) .
\end{equation}
For finite alphabets for inputs and outputs, the set of all these
correlations is convex with a finite number of extremal points,
hence it is a polytope. This new conceptual tool allows one for
the first time to study quantum nonlocality {\it from the
outside}, that is without all the Hilbert space machinery. Several
recent papers explore this new avenue
\cite{NSC,CGMP,BP,Barrett,us}. In particular, it is proven in
\cite{Barrett,us} that no-signaling alone implies that nonlocal
correlations are also monogamous since they cannot be cloned.

It is thus natural to ask whether, as suggested by Artur Ekert,
the security of QKD does not rely, ultimately, on quantum
nonlocality. At first sight this seems unlikely. %Indeed, the
%correlation obtained by Alice and Bob when running the best known
%BB84 QKD protocol do not violate any Bell inequality, i.e. the
%BB84 correlation can be simulated by local variables. But then,
%why can't Eve hold a copy of these local variables?
The standard answer runs as follows: if Alice and Bob are
sufficiently entangled, then the adversary Eve is essentially
factorized out. Despite this standard answer, the situation is
more subtle: If the Alice-Bob correlation is local, it can be
reproduced by (classical or quantum) variables coming from a
source. However, a perfect copy of these variables could also be
sent to Eve. Then it could be that Alice and Bob share a separable
states of larger dimension! Indeed, the local variables
necessarily enlarge Alice and Bob's Hilbert space. As an example,
consider BB84 \cite{BB84}, where Alice and Bob choose between two
different measurements. Their correlation can be reproduced by the
four-qubit separable state
\begin{equation}
    \rho_{AB}=\frac{1}{4}(\proj{00}_z+\proj{11}_z)\otimes
    (\proj{00}_x+\proj{11}_x) .
\end{equation}
Here, Alice holds the first and third qubit. Whenever she measures
in the $z$ ($x$) basis she is actually measuring the first (third)
qubit in this basis. The same happens for Bob, with the second and
fourth qubit. Clearly, their measurement results are completely
correlated when the bases agree and uncorrelated otherwise.
However their state is separable, so BB84 becomes insecure even in
the ideal noise-free situation! In summary, all security proofs of
QKD assume that the legitimate partners, Alice and Bob, know the
dimensions of the Hilbert space describing their quantum systems.
In practice, this is usually a reasonable assumption, however it
underlines that assumptions are necessary for any security proof.
Moreover it is conceptually interesting to {\it disentangle} the
consequences of no-signaling from those relying on the Hilbert
space formalism. Experimentally, additional Hilbert-space
dimensions correspond to ``side-channels", i.e. to degrees of
freedom coded accidentally. For example, in photon polarization
coding, the wavelength could be accidentally correlated to the
state of polarization.

In this letter, we first present a new 4-state QKD protocol, next
demonstrate its security against any individual attack by any
adversary only limited by the no-signaling condition \cite{BHK}.
In particular, Eve could be {\it supra-quantum}, since there are
nonsignaling correlations not achievable using quantum states
\cite{PR}. In the new protocol, Alice and Bob have for each
realization the choice between two measurements with binary
outcomes. Essential is that their measured data violate the
Clauser-Horne-Shimony-Holt (CHSH) Bell inequality \cite{CHSH}:
\begin{equation}\label{chsh}
    P(a_0=b_0)+P(a_0=b_1)+P(a_1=b_0)+P(a_1\neq b_1)\leq 3 ,
\end{equation}
where $P(a_j=b_k)=P(a=b=0|x=j,y=k)+P(a=b=1|x=j,y=k)$. For example,
Alice and Bob could share a Werner state,
$\rho_W=WP_{\phi^+}+(1-W)\opone/4$, with visibility $W$, where
$P_{\phi^+}$ denotes the projector onto
$\ket{\phi^+}=(\ket{00}+\ket{11})/\sqrt 2$, and perform the
measurements that maximize the violation of the CHSH-Bell
inequality (\ref{chsh}). But any other way to obtain data
violating (\ref{chsh}) is equally good. Hence we name our new
protocol the CHSH-protocol. Violation of (\ref{chsh}) implies that
in three out of the four measurement choices, Alice and Bob are
correlated (the three first terms in (\ref{chsh})), though not
necessarily maximally, while in the fourth case they are
anti-correlated. The analog of basis reconciliation goes as
follows: Bob announces all his measurement settings, Alice keeps
all her data, but for the case of
anti-correlation, she flips her bit. %After this reconciliation,
%their probability distribution becomes independent of the applied
%measurements.
Compared to BB84, %in this new protocol the raw key
%is twice as long, since
the partners keep all data, however, all data are noisy. %Note that
%the key length reduction necessary to correct for the noise
%($1-h(\frac{1-W}{2})$, where $h$ denotes the binary entropy
%function) is larger than $\half$.
For $W>1/\sqrt{2}$ the new protocol produces data that violate
Bell's inequality. Hence there are no local variables that Eve
could hold.

%\begin{figure}
  % Requires \usepackage{graphicx}
%  \includegraphics[width=8 cm]{encoding.eps}\\
%  \caption{Optimal measurements for CHSH inequality violation
%  by Werner states $\rho_W$.}\label{measurements}
%\end{figure}

In the sequence, we limit our analysis to isotropic raw
correlations with visibility $V$: \beq
P(a,b|x,y)=V\half\delta(a+b=x\cdot y) + (1-V)\frac{1}{4}
\label{isoCorrel} \eeq where $\delta(r=s)=1$ whenever the equality
holds modulo 2, and 0 overwise. For $V\le 1/\sqrt{2}$, such
correlations can be distributed by quantum physics (e.g. by a
Werner state with $W=\sqrt{2}V$) and for $V>1/2$ they violate the
Bell inequality (\ref{chsh}). This does not imply any loss of
generality: Alice and Bob can map any binary correlation into
these isotropic correlations by local operations and classical
communication keeping the Bell violation \cite{us}.

\begin{figure}
  % Requires \usepackage{graphicx}
  \includegraphics[width=6cm]{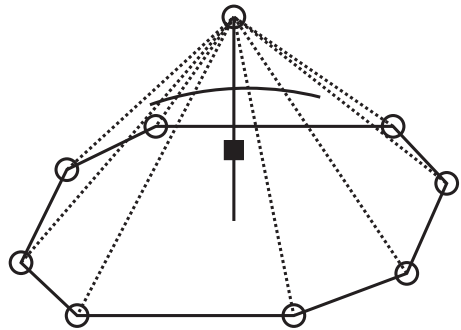}\\
  \caption{Pictorial representation of nonsignaling correlations
  $P(a,b|x,y)$ for binary inputs and outputs.
  The thick lines define one of the facets of the polytope of local
  models, or local polytope, corresponding to the CHSH inequality.
  All the extreme points lying on this facet are also extreme
  points of the more general polytope of nonsignaling
  correlations. Only one extremal point is on top
  of the CHSH facet, given by the nonlocal machine. The curved
  line schematically represents the region of points achievable
  using quantum states. Isotropic correlations (\ref{isoCorrel})
  lie along the vertical line starting from the nonlocal machine and entering
  the local polytope through the center of the CHSH facet.  In the
  optimal eavesdropping attack, Eve
  simulates Alice and Bob's correlation, square point, by
  the proper combination of extreme points of the general polytope,
  circles in the figure.
  }\label{polytope}
\end{figure}

Let us study the security of this protocol. As usual, it is
assumed that the distribution of the correlation is done by Eve.
Any attack consists of a three-party distribution $P(a,b,e|x,y,z)$
whose marginal is (\ref{isoCorrel}):
\begin{eqnarray}\label{eveattack}
    P(a,b|x,y)&=&\sum_e P(a,b,e|x,y,z) \nonumber\\
    &=&\sum_e P(e|z)P(a,b|x,y,z,e) .
\end{eqnarray}
where for the second equality we used the no-signaling condition:
Eve's output $e$ is independent of Alice and Bob's inputs $x$ and
$y$. We can restrict our considerations to attacks where Eve
prepares extreme points of Alice and Bob's no-signaling polytope.
Indeed, consider an attack where this is not the case, that is,
some of the terms appearing in Eq. (\ref{eveattack}) do not
correspond to an extreme point of Alice-Bob's no-signaling
polytope. Then, these terms can be expressed as a convex
combination of extreme points
\begin{equation}\label{eveconvex}
    P(a,b|x,y,z,e)=\sum_\lambda
    P_{\mbox{\small{ext}}}(a,b|x,y,z,e,\lambda) P(\lambda) .
\end{equation}
Giving the knowledge of $\lambda$ to Eve, one has an attack
consisting of extreme points and that is at least as good as the
previous one, since, c.f. (\ref{eveattack}),
\begin{equation}\label{eveextr}
    P(a,b|x,y)=\sum_{e,\lambda} P(e,\lambda|z)P_{\mbox{\small{ext}}}(a,b|x,y,z,e,\lambda) .
\end{equation}
%In fact, Eve can always transform the attack consisting of extreme
%points into the original one by forgetting $\lambda$.
We need to recall now some facts about nonsignaling correlations
with binary input and output. Barrett and co-workers proved that
in this simple binary case, the number of extremal nonsignaling
correlations is very limited \cite{NSC}. If moreover one
concentrates on the correlations that violate the CHSH-Bell
inequality, then one finds a unique extremal correlation that
violates it; this is the isotropic correlation (\ref{isoCorrel})
with $V=1$. This correlation appears in the literature as PR-box
\cite{NSC}, or NonLocal Machine \cite{CGMP} or Unit of Nonlocality
\cite{BP}. Moreover, there are 8 extremal correlations that
saturate the inequality (\ref{chsh}), see Fig. \ref{polytope}.

%We also restrict to the case where the noise observed by Alice and
%Bob is isotropic, i.e. equally distributed for all combination of
%settings. These are the points along the line in Figure
%\ref{polytope}, starting from the nonlocal machine and entering
%the quantum region and then the local polytope through the center
%of the CHSH facet. These correlations can be written as
%$a+b=xy+\xi$, where $\xi$ is a random binary variable independent
%of the
%settings. %such that $P(\xi=0)=\delta$. The nonlocal machine
%corresponds to $\delta=1$, the Tsirelson's bound to
%$\delta=(1+1/\sqrt 2)/2$, while the local point is $\delta=3/4$.
%The value of the CHSH inequality is $\beta=4\delta$.

For the local points, all the outcomes are deterministic and known
by Eve. However, if Alice and Bob share a nonlocal machine, they
have the guarantee of perfect monogamy \cite{NSC}, so Eve cannot
be correlated at all. Eve's optimal attack then consists of the
combination of extreme points that mimics Alice-Bob's correlation
with the minimal weight for nonlocal points. Therefore, she
prepares only those local points that are closer to Alice and
Bob's correlation. This can be easily understood in Figure
\ref{polytope}: in order to reproduce the quantum correlation
observed by Alice and Bob, represented by a square, Eve should
send an equal mixture of the eight local points lying on the
facet, plus the nonlocal machine on top of it. In what follows
$p_{NL}$ denotes the probability that Eve prepares a nonlocal
machine. The Bell violation observed by Alice and Bob fixes the
value of $p_{NL}$, since $p_{NL}=2V-1$. When the observed data are
local, Eve can mimic them with deterministic local points.
However, when the correlation is nonlocal, Eve is sometimes forced
to send a nonlocal machine, where she cannot be correlated because
of the no-signaling principle. The resulting probability
distribution, after basis reconciliation, is summarized in Table
\ref{evetable}, where $p_L=1-p_{NL}$. Eve's information on Alice
and Bob's outcomes is represented by two variables $(e_a,e_b)$.
The value at each position of the table gives the probability for
the corresponding outcomes, e.g. $P(a=0,b=0,e=(?,0))=p_L/8$.
Notice that since only Bob announces his measurement, Eve
sometimes has deterministic information on Bob's but not on
Alice's symbol after the basis reconciliation, even if her
preparation was local. For example, consider the instance when Bob
announces $y=1$, then, even when Eve knows $a_0$ and $a_1$, she
might not know Alice's output. Moreover, one can see that, due to
the properties of the local points lying on the CHSH facet, Eve
has full information on both outcomes only when $a=b$.

\begin{table}
  \centering
  \begin{tabular}{|c|c|c|}
  % after \\: \hline or \cline{col1-col2} \cline{col3-col4} ...
  \hline
  \ \ $b$ & 0 & 1 \\
  $a$ $(e)$ &   &   \\
  \hline
   & (0,0) $p_L/4$ &  \\
  0 & (?,0) $p_L/8$ & (?,0) $p_L/8$ \\
   & (?,?) $p_{NL}/2$ &   \\
   \hline
   &  & (1,1) $p_L/4$ \\
  1 & (?,1) $p_L/8$ & (?,1) $p_L/8$ \\
   &  & (?,?) $p_{NL}/2$ \\
   \hline
\end{tabular}
  \caption{Eve's optimal individual attack. Alice and Bob's variables are
  binary, while Eve's information can be represented by two ternary variables,
  $e_a,e_b=0,1,$ ?. The value '?' denotes those cases where she has no information.
  For example, (?,0) means that Eve knows $b=0$ but not $a$.}
  \label{evetable}
\end{table}

Once the optimal individual attack has been determined, it is time
to study the secrecy properties of the resulting probability
distribution. One can see that: (i) a secure key can already be
established with one-way communication protocols and quantum
states, (ii) this probability distribution contains secret
correlations if and only if the CHSH inequality is violated, i.e.
$p_{NL}>0$, and (iii) it seems challenging to reach the Bell
violation limit, $p_{NL}=0$, by means of the known two-way
advantage distillation techniques. The detailed calculation of
these results will be given in a forthcoming paper \cite{prepr}.

In the case of one-way distillation protocols, it is clear that
the flow of information has to go from Alice to Bob. Indeed, since
Bob
announces the basis, Eve's information on his outcome is larger%,
%$I(B:E)>I(A:E)$
. From Table \ref{evetable} one has that Bob's
error probability is $\varepsilon_{B}=p_L/4$, while
$I(A:E)=p_L/2$. Then, the one-way key rate, $K^\rightarrow$,
satisfies \cite{CK}
\begin{equation}\label{ckbound}
    K^\rightarrow\geq I(A:B)-I(A:E)=1-h(p_L/4)-\frac{p_L}{2} ,
\end{equation}
where $h$ is the binary entropy. This quantity is positive for
$p_{NL}\gtrsim 0.318$. The quantum region is given by $p_{NL}\leq
\sqrt 2-1\simeq 0.414$, so quantum correlations suffice to
achieve security against individual nonsignaling attacks. %Notice
%that if the reconciliation was from Bob to Alice, Eve's
%information would be $I(B:E)=p_L$ and the corresponding security
%bound would change to $p_{NL}\simeq 0.525$, not achievable by
%quantum states!

Next, one can prove that the obtained probability distribution
contains secret correlations if and only if there is a Bell
inequality violation. We take as a measure of secret correlations
the so-called intrinsic information, $I(A:B\downarrow E)$ or more
briefly $I_\downarrow$, introduced in \cite{MW}. A tripartite
probability distribution can be established by public
communication if and only if the intrinsic information is zero
\cite{RW}. It was shown in \cite{us} that if none of the parties
announces the choice of bases, the set of local and public
correlations are equivalent, under the no-signaling principle.
Here, Bob announces his basis through the public channel, so it
could happen that the honest parties loose some secrecy. One can
see however, that the intrinsic information still remains positive
for the whole region of Bell violation, since
\begin{equation}\label{intrinf}
    I_\downarrow= h(1-p_{NL}/2) - \frac{1+p_{NL}}{4}\,
    h\left(\frac{1-p_{NL}}{1+p_{NL}}\right) .
\end{equation}
In order to get this result, we numerically compute $I_\downarrow$
for different values of $p_{NL}$. In all the cases, we found a
perfect agreement with this formula. Interestingly, if Alice
announces her basis too, $I_\downarrow=0$ when
$p_{NL}\leq1/5$~\cite{prepr}.

We also analyze the use of the two-way advantage distillation
protocol of Ref. \cite{Maurer}. Advantage distillation
moves the region of positive key rate to $p_{NL}>1/5$ \cite{prepr}. %The proof
%of this result follows from~\cite{TH}.

\begin{figure}
  % Requires \usepackage{graphicx}
  \includegraphics[width=8cm]{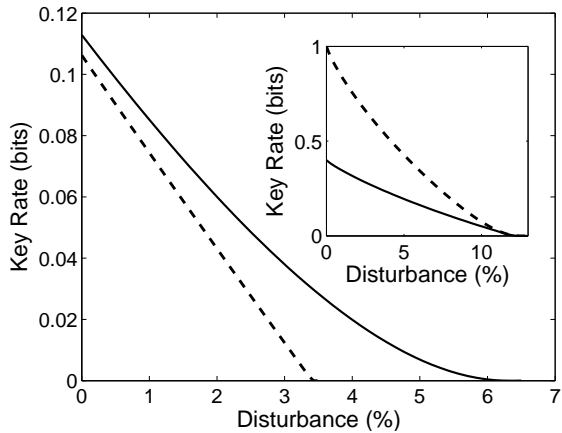}\\
  \caption{Key rates using one-way communication against
  nonsignaling individual attacks. The dashed line corresponds
  to the situation where no pre-processing is employed, while the
  thick one to the optimal pre-processing. In the inset, the key
  rate against a standard quantum eavesdropper is given, compared to BB84
  (dashed line).}\label{rates}
\end{figure}

The use of pre-processing by the parties, as studied in Ref.
\cite{KGR}, is useful in all the situations. %For example, Alice
%introduces some noise, flipping her outcome, $a\rightarrow 1-a$,
%with
%probability $\gamma$. %This worsens her correlations with Bob, but
%may deteriorate her correlations with Eve in a
%stronger way. %After this pre-processing, Bob's error probability
%and Eve's information change to
%\begin{eqnarray}
%\label{errpr}
% \nonumber to remove numbering (before each equation)
%  &&\varepsilon'_{B} = (1-\gamma)\varepsilon_{B}+
%  \gamma(1-\varepsilon_{B}) \nonumber\\
%    &&I'(A:E) = \frac{p_L}{2}(1-h(\gamma)) .
%\end{eqnarray}
The corresponding one-way key rate as a function of the
disturbance $D$ in the quantum channel is shown in Figure
\ref{rates}. The disturbance is defined in the standard way,
namely $D=0$ corresponds to a perfect channel, and $p_{NL}=\sqrt
2(1-2D)-1$. The critical disturbance is $D\lesssim 6.3\%$. Note a
fundamental difference between our security analysis and Ekert's
protocol: in
his scheme, %to observe the maximal quantum violation of the CHSH
%inequality,
$D=0$ guarantees perfect security. This is not the case here,
since the eavesdropper is only limited by the no-signaling
principle. In the case of two-way communication, a positive key
rate is obtained when $D\lesssim 11.36\%$, or $p_{NL}\gtrsim
0.093$, still not sufficient to cover the region of
Bell violation. %Indeed, in the case of qubits and a
%quantum eavesdropper, the entanglement limit can be shown to
%coincide with the security limit under individual attacks just
%using standard advantage distillation~\cite{AMG}. Moreover,
%Taking into account that Bell violation guarantees that the
%observed probability distribution contains secrecy,
Therefore, the binary probability distribution of
Table~\ref{evetable}, when the Bell violation is small, either
contains bipartite bound information~\cite{GW} or is distillable
using a new technique. Both alternatives appear very interesting.

Finally, it is also relevant to study the protocol in the standard
scenario where the eavesdropper is quantum and Alice and Bob know
their Hilbert spaces. A general security proof is beyond the scope
of the present work. Here, we study the simple case of collective
attacks, where Alice, Bob and Eve are assumed to share copies of
the same state. We compare the obtained rates with those for the
BB84 protocol \cite{BB84}
in Ref. \cite{KGR}, see Fig. \ref{rates}. %As one can see in Figure \ref{rates},
%the robustness of the protocol seems to be the same as for BB84.
%Recall that in the present protocol, Alice and Bob data are noisy
%even in the case of no disturbance, the error probability being
%$\sin^2(\pi/8)\simeq 0.146$. Indeed,
The quantum rates for our protocol without pre-processing are the
same as for BB84 with pre-processing given by
$\gamma=\sin^2(\pi/8)$ \cite{prcomm}. For high disturbances, Alice
starts adding noise, achieving the same rates and critical
disturbance as for BB84, namely $D=12.4\%$ \cite{KGR}.

To summarize, usual security proofs of QKD are based on
entanglement theory. However, entanglement theory, like e.g.
entanglement witnesses, assume fixed dimensions for Alice and
Bob's Hilbert spaces. If data do not violate any Bell inequality,
then they can be produced by quantum measurements on a separable
state of large dimension (the extra dimension allows one to
include all supplementary local variables). Hence no QKD could be
achieved without either a violation of some Bell inequality, or
assumptions about the dimension of the relevant Hilbert spaces. In
this letter we presented a QKD protocol aimed at producing data
that violate the CHSH-Bell inequality. We proved its security
against the most general individual attack without signaling,
independently of any assumption about Hilbert spaces. To our
knowledge, our results represent the first step towards the
characterization of optimal nonsignaling eavesdropping attacks.
Note also that the same data, i.e., same state preparation and
measurements, can be secure in the standard quantum scenario or
against no-signalling eavesdroppers: only the amount of error
correction and privacy amplification has to be changed. %Many
%generalizations of this result are worth studying: for instance,
%the use of more inputs or larger alphabets to increase the
%robustness, or the analysis of non-isotropic correlations.
We would like to conclude with a comment on the role played by
Bell inequalities in our discussion. It is often said that any
Bell inequality is just an example of an, often non-optimal,
entanglement witness. However, they are more than this, since they
are derived without invoking the quantum formalism. As shown here,
they are witnesses of useful correlations independent of the
Hilbert space structure.

%In this work, we present a quantum key distribution scheme and
%prove its security against individual nonsignaling attacks. The
%crucial ingredient for security comes from the violation of the
%CHSH inequality. Indeed, nonlocal correlations represent a
%resource for monogamy in general no-signaling theories \cite{us},
%in a similar way as entanglement in Quantum Mechanics. Note also
%that the same data, i.e., same states preparation and
%measurements, can be secure against quantum or no-signaling
%eavesdroppers. One only has to change the amount of error
%correction and privacy amplification. Actually, in the quantum
%scenario, our protocol is as robust as BB84 and for, high
%disturbances, gives the same key rates. However, it is also secure
%in the no-signaling scenario, where BB84 is completely insecure.

%There are several questions that follow from our work: for
%instance, it would be interesting to extend the security analysis
%to general no-signaling attacks and to compare this protocol with
%the one in \cite{BHK}.

%\section{Acknowledgements}

\bigskip

We thank S. Iblisdir, B. Kraus and V. Scarani for useful
discussion. This work is supported by a Spanish MCYT ``Ram\'on y
Cajal" grant, the Generalitat de Catalunya, the Swiss NCCR
``Quantum Photonics" and OFES within the EU project RESQ
(IST-2001-37559), and the U.K. Engineering and Physical Sciences
Research Council (IRC QIP).

\end{document}